\documentclass{article}

\usepackage{arxiv}

\usepackage[utf8]{inputenc} 
\usepackage[T1]{fontenc}    
\usepackage{hyperref}       
\usepackage{url}            
\usepackage{booktabs}       
\usepackage{amsfonts}       
\usepackage{nicefrac}       
\usepackage{microtype}      
\usepackage{lipsum}		
\usepackage{graphicx}
\usepackage{doi}
\usepackage{subcaption}

\title{Quantum Machine Learning: Unveiling Trends, Impacts through Bibliometric Analysis}

\author{ \href{https://orcid.org/0009-0009-8983-1436}{\includegraphics[scale=0.06]{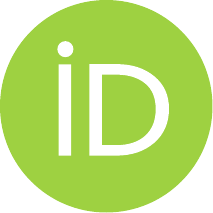}\hspace{1mm}Riya  Bansal} \\
	Department of Computer Science\\
	University of Delhi\\
	Delhi, India 110007 \\
	\texttt{rbansal1@cs.du.ac.in} \\
	\And
	\href{https://orcid.org/0000-0002-7268-0207}{\includegraphics[scale=0.06]{orcid.pdf}\hspace{1mm}Nikhil Kumar Rajput}\thanks{Corresponding author} \\
	Ramanujan College\\
	University of Delhi\\
	Delhi, India 110019 \\
	\texttt{n.rajput@ramanujan.du.ac.in} \\
	}
\renewcommand{\shorttitle}{\textit{arXiv} Template}

\hypersetup{
pdftitle={Quantum Machine Learning: Unveiling Trends, Impacts through Bibliometric Analysis},
pdfsubject={q-bio.NC, q-bio.QM},
pdfauthor={Riya Bansal, Nikhil Kumar Rajput},
pdfkeywords={Quantum Machine Learning, Bibliometric Analysis, Quantum computing, Machine learning},
}

\begin{document}
\maketitle

\begin{abstract}
	Quantum Machine Learning (QML) is the intersection of two revolutionary fields: quantum computing and machine learning. It promises to unlock unparalleled capabilities in data analysis, model building, and problem-solving by harnessing the unique properties of quantum mechanics. This research endeavors to conduct a comprehensive bibliometric analysis of scientific information pertaining to QML covering the period from 2000 to 2023. An extensive dataset comprising 9493 scholarly works is meticulously examined to unveil notable trends, impact factors, and funding patterns within the domain. Additionally, the study employs bibliometric mapping techniques to visually illustrate the network relationships among key countries, institutions, authors, patent citations and significant keywords in QML research. The analysis reveals a consistent growth in publications over the examined period. The findings highlight the United States and China as prominent contributors, exhibiting substantial publication and citation metrics. Notably, the study concludes that QML, as a research subject, is currently in a formative stage, characterized by robust scholarly activity and ongoing development.
\end{abstract}


\section{Introduction}

Machine learning has been the driving force behind the remarkable evolution of artificial intelligence, enabling computers to learn from data, extract patterns, and make predictions (\cite{bi2019machine}). While classical algorithms like regression and deep learning have led to breakthroughs in areas like image recognition and predictive modeling, the ever-growing complexity of real-world problems and escalating computational demands are pushing the limits of conventional computing. This necessitates a paradigm shift, and the emergence of Quantum Machine Learning (QML) (\cite{biamonte2017quantum, wittek2014quantum, schuld2015introduction}) presents a promising solution.
QML blends two powerful domains: the enigmatic world of quantum mechanics, governed by principles like superposition and entanglement, and the data-driven power of machine learning. Quantum mechanics (\cite{messiah2014quantum}) unveils the bizarre behavior of atoms and subatomic particles, where they can exist in multiple states simultaneously and exhibit unique connections. In contrast, machine learning empowers computers to autonomously learn from data, paving the way for technologies like recommendation systems.

The union of these two seemingly disparate fields, QML, has ignited tremendous excitement and holds immense potential (\cite{schuld2014quest, dunjko2018machine}). It envisions integrating quantum algorithms into machine learning programs, leveraging the capabilities of quantum devices like quantum computers (\cite{ladd2010quantum}) to complement, accelerate, or enhance the performance of classical algorithms. Commonly referred to as "quantum-enhanced machine learning" (\cite{dunjko2016quantum}), QML harnesses the extraordinary information processing capabilities of quantum technology to amplify and expedite machine learning operations. While classical algorithms struggle with vast amounts of data, QML utilizes qubits and specialized quantum systems to revolutionize computational speed and data storage (\cite{yoo2014quantum, khan2020machine}). This transformative potential lies in QML's ability to offer exponential speedups and enhanced accuracy in data processing and analysis, surpassing the limitations of classical methods. However, hurdles remain before QML's full potential can be unleashed. Hardware limitations and the challenge of training quantum machine learning models pose significant obstacles (\cite{cerezo2022challenges, ciliberto2018quantum, houssein2022machine}). Overcoming these challenges will be key to unlocking the transformative power of QML and ushering in a new era of AI, fueled by the unique capabilities of quantum computing.

Recent attention in quantum technology research, as reflected in bibliometric analysis (\cite{scheidsteger2021bibliometric}), underscores the increasing importance of understanding the research landscape of QML. This interdisciplinary field has garnered substantial interest, particularly since 2014, as evidenced by its promising role in Noisy Intermediate-Scale Quantum (NISQ) computer (\cite{cheng2023noisy}) research. A notable study conducted by \cite{pande2020bibliometric} specifically delves into the quantitative assessment of QML research. \cite{dhawan2021quantum} conducted a comprehensive bibliometric survey using the Scopus database to examine publications in QML spanning a 22-year period from 1999 to 2020. This research delved into global research output and citations, employing metrics to assess productivity and performance. Additionally, the study employed bibliometric mapping techniques to visually depict the network relationships among key countries, institutions, authors, and significant keywords in the field of QML research. A Bibliometric Analysis and Visualization of Quantum Engineering Technology was done by \cite{sood2023bibliometric}. The results underscore that quantum algorithms form the foundational cornerstone across various research fields, with QML emerging as the most dynamically active area of research in recent times. The study by  \cite{wang2021bibliometric} endeavors to provide a thorough and detailed analysis of the global development status of quantum computing. Wang suggests that quantum computing is positioned as the next-generation technology crucial for the future of human society. 

In essence, this study collectively signify a growing interest in comprehending the intricate landscape of QML through bibliometric analysis. In this work, we are reviewing the literature on quantum machine learning and provides the statistical analysis of the articles present in the Lens database. This analysis involves publication and citation analysis, which field contributed the most in research, most productive countries and institutions. It also analyses most productive authors, highly cited articles, most productive journals, significant keywords uses in the articles, patent citations and patent cited work and, the top organizations that funded for the research to conduct. This approach proves crucial in gaining insights into the development, challenges, and impact of this dynamic and evolving field.

The remainder of the study is organised as follows: Section \ref{methodology} tells about the methodology adopted to gather the articles for analysis. Section \ref{results} discusses the observations/results obtained after the analysis of the articles. Section \ref{conclusion} concludes the study while the limitations of the study are mentioned in section \ref{limitations}.

\section{Methodology} \label{methodology}

\begin{figure}
    \centering
    \includegraphics[width=0.8\linewidth]{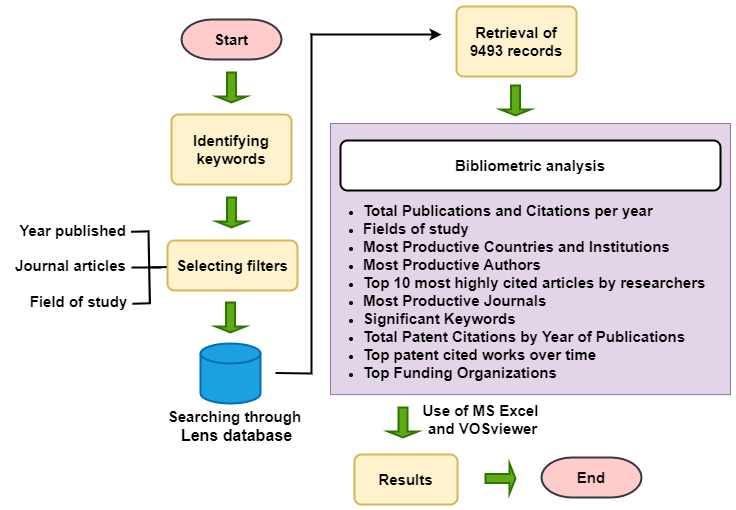}
    \caption{Workflow of bibliometric analysis}
    \label{fig:methodology}
\end{figure}

To comprehensively explore the publishing trends and patterns in QML, bibliometric data was systematically gathered from the Lens database \url{https://www.lens.org/}. A focused search was conducted using the keyword ``quantum machine learning" in the ``search by keyword or scholarly field" feature. The search was refined through the application of specific filters to ensure precision and relevance.

The applied filters encompassed the following criteria:

\textbf{Filters:} 
Year Published = ( 2000 - 2023  )
Publication Type = ( journal article  ) 
Field of Study = ( Computer science  , Quantum mechanics  , Machine learning  , Quantum computer  , Quantum machine learning  , Quantum algorithm  , Quantum  )

This filtering process resulted in the retrieval of 9493 records. These records, encapsulating the scholarly output within the specified parameters, were then downloaded in CSV file format for subsequent in-depth analysis. The analytical phase involved the utilization of various tools, including MS Excel and VOSviewer, to unravel and interpret the intricate bibliometric patterns and insights inherent in the vast dataset. The whole process of finding out the records and conducting the bibliometric analysis is visualized in Figure \ref{fig:methodology}.

\section{Results} \label{results}

This section deals with the articles that were retrieved from the database and provides a bibliometric analysis for the same. The obtained articles were analysed for publications and citations per year, field of study, most productive countries, institutions, authors and journals, most significant keywords, patent analysis and, contribution of funding organizations. 

\subsection{Total Publications and Citations per year}

\begin{figure*}[!ht]
    \centering
    \includegraphics[width=0.8\linewidth]{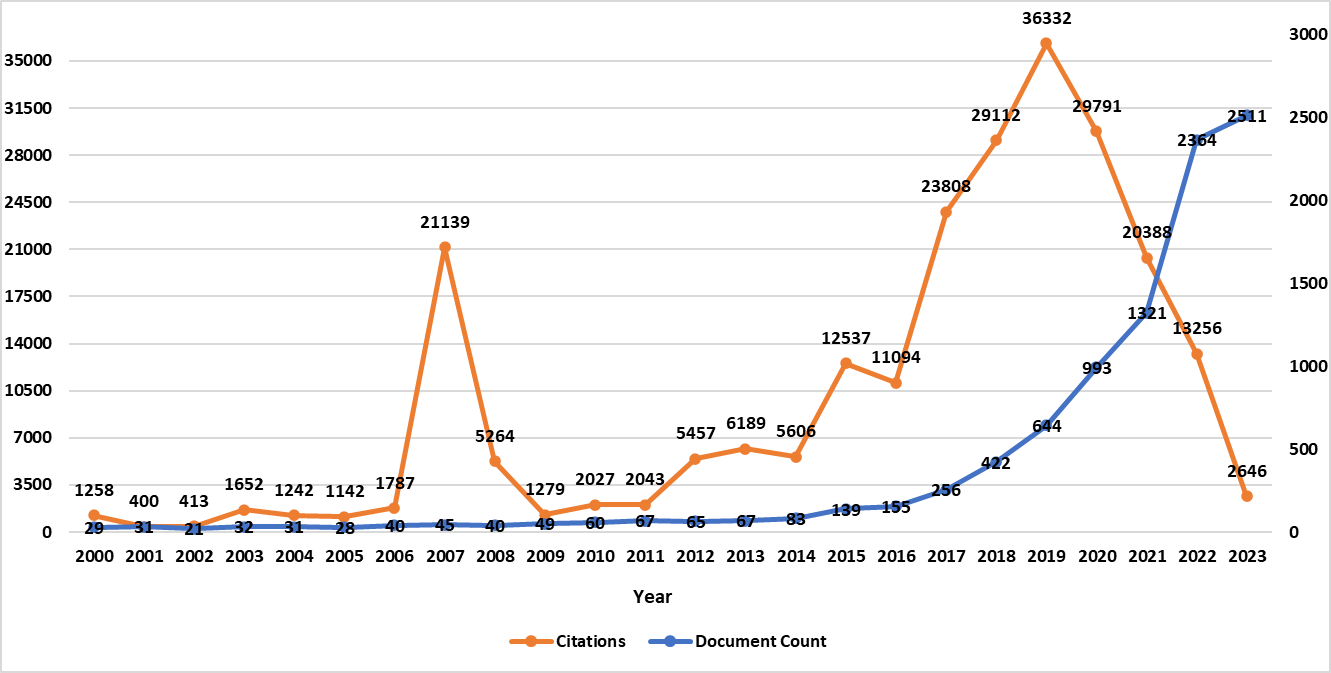}
    \caption{Total Publications and citations per year}
    \label{fig:publications-citations-per-year}
\end{figure*}

The visual representation in Figure \ref{fig:publications-citations-per-year} provides a comprehensive overview of both the annual count of published scholarly works and the corresponding scholarly citations, totaling 9493 and 235862 respectively, between the years 2000 and 2023. In the initial years, spanning from 2000 to 2009, there was a relatively modest volume of publications. To be specific, the annual publication count was below 50 in the initial nine years starting from 2000. This figure distinctly highlights a notable trend within the field, signaling a surge in publications that gained momentum starting from the year 2015. The data reveals a significant increase in scholarly output, with the number of publications in 2023 nearly eighteen times that of the publications in 2015. This observed acceleration in research output suggests substantial growth and heightened interest in the field over the specified timeframe.

Furthermore, it depicts a distinct pattern in scholarly citations. Notably, there is a sudden increase in the number of citations in the year 2007, followed by a subsequent drop from 2008 onwards. The trend then experiences a resurgence from the year 2017, and decrease again from the year 2020. These fluctuations in scholarly citations may be indicative of pivotal moments or breakthroughs in the field, warranting closer examination to understand the dynamics influencing the scholarly impact over different periods.

\subsection{Fields of study}

\begin{figure}[!ht]
    \centering
    \includegraphics[width=0.8\linewidth]{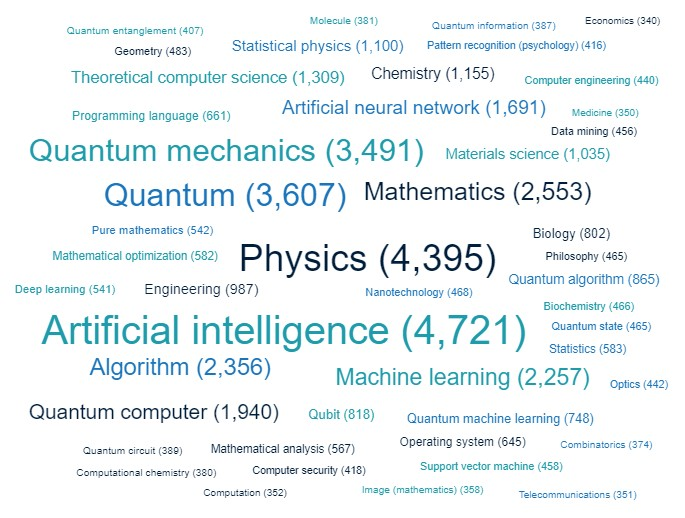}
    \caption{Fields of study}
    \label{fig:field-of-study}
\end{figure}

The visual representation in Figure \ref{fig:field-of-study} offers a word cloud depicting the various fields contributing to the research in question. Among the 9493 articles analyzed, the field of Artificial Intelligence (4721) emerges as the predominant contributor, constituting over more than 50\% of the total published articles. Additionally, significant contributions are observed from fields such as Physics (4395), Quantum (3607), Quantum Mechanics (3491), and others, collectively forming a diverse landscape of research interests. This visualization effectively highlights the prominence of Artificial Intelligence and showcases the multifaceted nature of the research, where various disciplines converge to contribute to the broader field.

\subsection{Most Productive Countries and Institutions}

The examination of research institutions and countries provides valuable insights into the leading contributors to this field of study. In Table \ref{tab:institution-countries}, the top 20 most influential countries and institutions are highlighted based on the total number of published articles. Just four institutions generated more than 100 publications. Leading the list is the Chinese Academy of Sciences, which has published a substantial 176 articles, closely followed by Tsinghua University with 160 articles and the Massachusetts Institute of Technology with 146 publications. While the Chinese Academy of Sciences boasts the highest number of publications, it is noteworthy that the Max Planck Society demonstrates significant impact and contribution, evidenced by an impressive 14055 total citations.

Among the top 20 countries, United States takes the lead with the highest number of published articles, totaling 2045. Following closely are China and United Kingdom with highest number of articles. These findings shed light on the global landscape of quantum machine learning research, highlighting specific institutions and countries that play a prominent role in shaping this dynamic field.  

\begin{table*}[]
\centering
\caption{Top 20 most influential institution and countries}
\scriptsize
  \begin{tabular}{llcccccc}
  \toprule
\textbf{Rank} & \textbf{Institution}                            & \textbf{Document Count} & \textbf{Total citation} & \textbf{Rank} & \textbf{Country} & \textbf{Document Count} & \textbf{Total citation} \\ 
\midrule
1             & Chinese Academy of Sciences                     & 176                                                                     & 3692                                                                    & 1              & United States    & 2045                                                                     & 125518                                                                  \\ 
2             & Tsinghua   University                           & 160                                                                     & 3527                                                                    & 2              & China            & 1781                                                                     & 38024                                                                   \\ 
3             & Massachusetts Institute of   Technology         & 146                                                                     & 12726                                                                   & 3              & United Kingdom   & 667                                                                      & 30301                                                                   \\ 
4             & University   of Science and Technology of China & 103                                                                     & 2541                                                                    & 4              & Germany          & 543                                                                      & 36142                                                                   \\ 
5             & Harvard University                              & 88                                                                      & 4195                                                                    & 5              & India            & 489                                                                      & 4901                                                                    \\ 
6             & Russian   Academy of Sciences                   & 84                                                                      & 666                                                                     & 6              & Japan            & 354                                                                      & 10377                                                                   \\ 
7             & ETH Zurich                                      & 83                                                                      & 5541                                                                    & 7              & Italy            & 305                                                                      & 7710                                                                    \\ 
8             & Peking   University                             & 83                                                                      & 3982                                                                    & 8              & Switzerland      & 290                                                                      & 13870                                                                   \\ 
9             & University of Oxford                            & 82                                                                      & 5884                                                                    & 9              & Canada           & 289                                                                      & 13634                                                                   \\ 
10            & University   of Technology, Sydney              & 80                                                                      & 6878                                                                    & 10             & Australia        & 277                                                                      & 12351                                                                   \\ 
11            & Max Planck Society                              & 78                                                                      & 14055                                                                   & 11             & Korea            & 244                                                                      & 11619                                                                   \\ 
12            & University   of Chinese Academy of Sciences     & 77                                                                      & 1443                                                                    & 12             & Spain            & 214                                                                      & 4008                                                                    \\ 
13            & Istituto Nazionale di Fisica   Nucleare         & 76                                                                      & 927                                                                     & 13             & France           & 206                                                                      & 8034                                                                    \\ 
14            & Oak   Ridge National Laboratory                 & 72                                                                      & 7185                                                                    & 14             & Russia           & 205                                                                      & 5661                                                                    \\ 
15            & Argonne National Laboratory                     & 70                                                                      & 4459                                                                    & 15             & Netherlands      & 156                                                                      & 4898                                                                    \\ 
16            & Shanghai   Jiao Tong University                 & 70                                                                      & 1313                                                                    & 16             & Singapore        & 130                                                                      & 4817                                                                    \\ 
17            & National University of Singapore                & 69                                                                      & 2408                                                                    & 17             & Iran             & 127                                                                      & 1650                                                                    \\ 
18            & Zhejiang   University                           & 69                                                                      & 1606                                                                    & 18             & Austria          & 121                                                                      & 5690                                                                    \\ 
19            & Centre national de la recherche   scientifique  & 65                                                                      & 1849                                                                    & 19             & Poland           & 109                                                                      & 3088                                                                    \\ 
20            & Purdue University                               & 64                                                                      & 2188                                                                    & 20             & Saudi Arabia     & 93                                                                       & 1358                                                                    \\ 
\bottomrule
\end{tabular}%
\label{tab:institution-countries}%
\end{table*}%

\subsection{Most Productive Authors}

\begin{figure}[h!]
    \centering
    \begin{subcaptionbox}{Top 20 authors by total documents\label{fig:documents}}%
        {\includegraphics[width=0.48\textwidth]{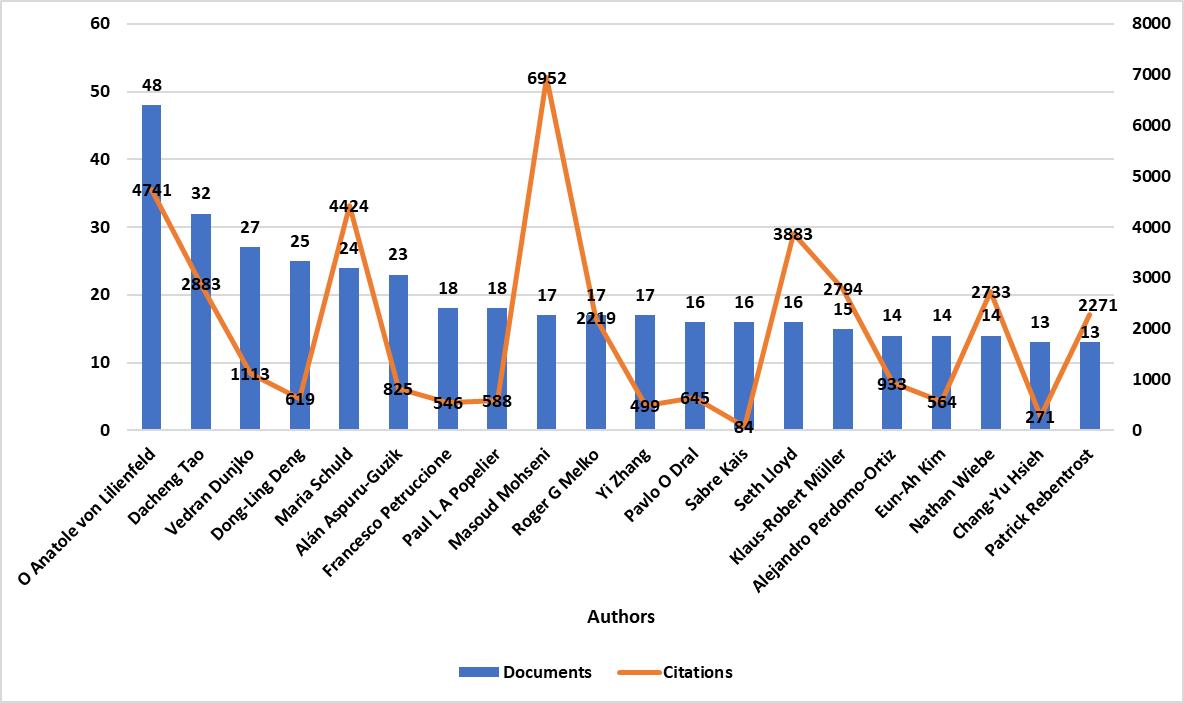}}
    \end{subcaptionbox}
    \hfill
    \begin{subcaptionbox}{Top 20 authors by cumulative citations\label{fig:citations}}%
        {\includegraphics[width=0.48\textwidth]{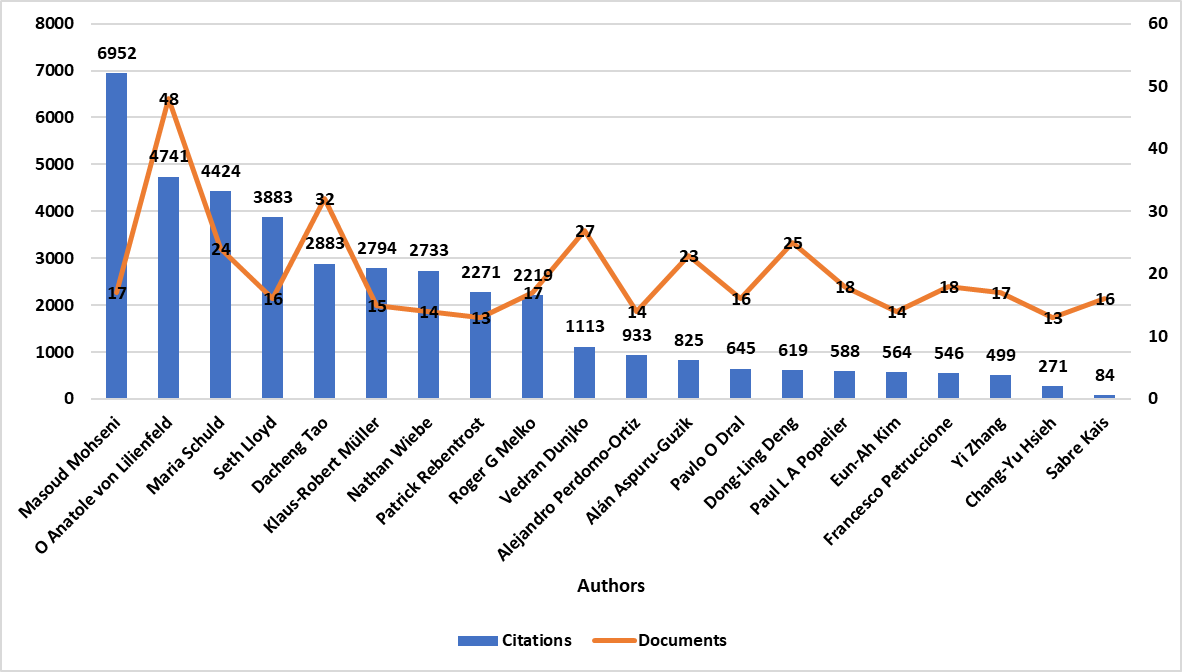}}
    \end{subcaptionbox}
    \caption{Top 20 most productive authors ranked by (a) number of documents and (b) cumulative citations.}
    \label{fig:authors}
\end{figure}

In the proposed research field, a cohort of 100 authors has been examined, and their impact is visualized in Figure \ref{fig:authors}. This figure delineates the top 20 most influential authors based on two key metrics: the number of total documents authored (Figure \ref{fig:documents} and the cumulative citations received (Figure \ref{fig:citations} over the period spanning from 2000 to 2023.

In terms of document production (Figure \ref{fig:documents}, O Anatole von Lilienfeld takes the lead among the top 20 authors, contributing to 48 documents. Following closely are Dacheng Tao and Vedran Dunjko, with 32 and 27 articles published, respectively. This metric provides insights into the prolificacy and contribution of authors in generating scientific literature within the specified timeframe. Contrastingly, Figure \ref{fig:citations} presents a distinct perspective, ranking authors based on the total number of citations received. Here, O Anatole von Lilienfeld appears in the second position with 4741 citations, emphasizing the significance of his work in garnering academic recognition. Topping the list in terms of citations is Masoud Mohseni, recognized as the most cited author with a remarkable 6952 citations. This dual perspective offers a nuanced understanding of author impact, considering both the volume and influence of their contributions to the field of study.

\subsection{Top 10 most highly cited articles by researchers}

 Table \ref{tab:cited-articles} provides a compilation of the top 10 highly cited articles including year, source, author, total citations and citing patents on Quantum Machine Learning. Summarizing the initial top two works, the article titled The quantum internet by \cite{kimble2008quantum}, published in the year 2008, claims the top spot with an impressive 4351 citations and 17 citing patents. The study tells that the quantum networks present both opportunities and challenges in the realms of quantum computation, communication, and metrology. The research conducted by \cite{arute2019quantum} highlights the exponential speed advantage of quantum computers over classical counterparts. Using a processor with programmable superconducting qubits, the study had achieved a groundbreaking feat by generating quantum states on 53 qubits, expanding the computational state-space significantly and signifies a major paradigm shift in computing. Article titled Quantum Machine Learning by \cite{biamonte2017quantum} stands out for having the highest number of citing patents 
 (18) among the listed articles. It is interesting that the majority of these influential articles find their home in the prestigious journal Nature, boasting citations exceeding 2000.

\begin{table*}[]
\centering
\caption{Top 10 most highly cited articles by researchers}
\scriptsize
  \begin{tabular}{lp{6cm}cp{2cm}p{3.5cm}p{1cm}p{1cm}}
  \toprule
\textbf{} & \textbf{Title}            & \textbf{Year} & \textbf{Source}                & \textbf{Author}                                                                                   & \textbf{Total Citations} & \textbf{Citing Patents} \\ 
\midrule
1 & The quantum internet                         & 2008          & Nature                         & Kimble, H. Jeff.                                                                                  & 4351                                                                & 17                                                                 \\ 
2 & Quantum supremacy using a programmable superconducting processor                                         & 2019          & Nature                         & Arute, Frank, et al.                                                                              & 4321                                                                & 0                                                                  \\ 
3 & Machine learning for molecular and materials science.                                                                                                                   & 2018          & Nature                         & Butler, Keith T., et al.                                                                          & 2304                                                                & 6                                                                  \\ 
4 & Quantum machine learning                                                                                                                                                & 2017          & Nature                         & Biamonte, Jacob, et al.                                                                           & 2023                                                                & 18                                                                 \\ 
5 & A comprehensive survey: artificial bee colony (ABC) algorithm and applications                              & 2012          & Artificial Intelligence Review & Karaboga, Dervis, et al.                                                                          & 1582                                                                & 3                                                                  \\ 
6 & Materials Design and Discovery with High-Throughput Density Functional Theory: The Open Quantum Materials Database (OQMD) & 2013          & JOM                            & Saal, James E., et al.                                                                            & 1419                                                                & 11                                                                 \\ 
7 & MoleculeNet: a benchmark for molecular machine learning                                                                   & 2017          & Chemical Science               & Wu, Zhenqin, et al.                                                                               & 1351                                                                & 14                                                                 \\ 
8 & Solving the quantum many-body problem with artificial neural networks                                   & 2017          & Science (New York, N.Y.)       & Carleo, Giuseppe, and Matthias Troyer                & 1328                                                                & 3                                                                  \\ 
9 & Machine learning and the physical sciences                                                                                                                              & 2019          & Reviews of Modern Physics      & Carleo, Giuseppe, et al.                                                                          & 1253                                                                & 0                                                                  \\ 
10 & Quantum Support Vector Machine for Big Data Classification                                                             & 2014          & Physical review letters        & Rebentrost, Patrick, Masoud Mohseni, and Seth Lloyd & 1066                                                                & 16                                                                 \\ 
\bottomrule
\end{tabular}%
\label{tab:cited-articles}%
\end{table*}%

\subsection{Most Productive Journals}

Scholarly journals provide an extensive repository of academic articles, serving as valuable resources for comprehending the evolution of knowledge within a research domain and generating insights for future investigations. 

Table \ref{tab:journal} presents the top 10 academic journals that have published major articles related to Quantum Machine Learning and had played a pivotal role in fostering communication among scholars, acting as a foundation for the cultivation of new ideas, and monitoring emerging trends and concepts. Notably, Scientific Reports leads the list with a maximum number of total publications in this field and have a decent impact factor of 4.6 (2022), followed by Quantum Information Processing and Nature Communications with 227 and 155 publications, respectively with an impact factor of 2.5 (2022) and 16.6 (2022), respectively. It is observed that the journal (Nature) with the highest impact factor of 64.8 (2022), having published 72 articles and have received the maximum citations (22287), which indicates its high prestige and reach within the field is positioned last among the top 10 journals . In terms of impact factors, it is noteworthy that the majority of the selected journals boast an impact factor exceeding 2, signifying their substantial academic influence and prominence within the scholarly community.

\begin{table*}[]
\centering
\caption{Most Productive Journals}
\scriptsize
  \begin{tabular}{llccc}
  \toprule
\textbf{} &
\textbf{Journal}                             & \textbf{Document Count} & \textbf{Total Citations} & \textbf{Impact Factor (2022)} \\ 
\midrule
1 & Scientific   reports                         & 329                                                                     & 7393                                                                     & 4.6                                                                           \\ 
2 & Quantum Information Processing               & 227                                                                     & 3161                                                                     & 2.5                                                                           \\ 
3 & Nature   communications                      & 155                                                                     & 10648                                                                    & 16.6                                                                          \\ 
4 & Physical Review A                            & 148                                                                     & 5066                                                                     & 2.9                                                                           \\ 
5 & Journal   of High Energy Physics             & 125                                                                     & 1640                                                                     & 5.4                                                                           \\ 
6 & npj Computational Materials                  & 119                                                                     & 7322                                                                     & 9.7                                                                           \\ 
7 & Machine   Learning-Science and Technology    & 99                                                                      & 1155                                                                     & 6.8                                                                           \\ 
8 & npj Quantum Information                      & 97                                                                      & 4657                                                                     & 7.6                                                                           \\ 
9 & Journal   of Chemical Theory and Computation & 91                                                                      & 6565                                                                     & 5.5                                                                           \\ 
10 & Nature                                       & 72                                                                      & 22287                                                                    & 64.8                                                                          \\ 
\bottomrule
\end{tabular}%
\label{tab:journal}%
\end{table*}%

\subsection{Significant Keywords}

\begin{figure*}[!ht]
    \centering
    \includegraphics[width=1\textwidth]{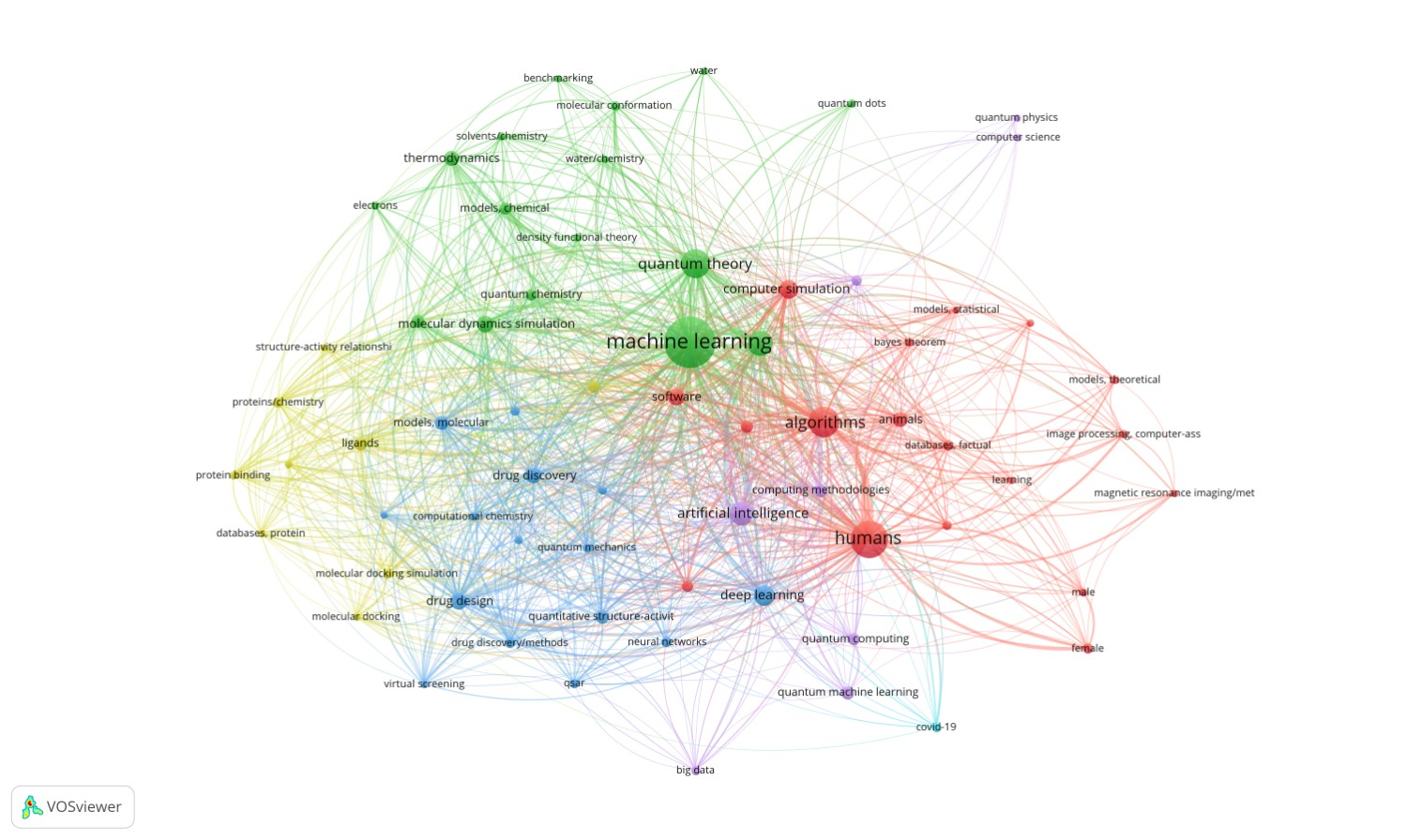}
    \caption{Keyword co-occurrence network visualization}
    \label{fig:co-occurence-keywords}
\end{figure*}

In Figure \ref{fig:co-occurence-keywords}, a comprehensive analysis of the global quantum machine learning literature yielded 67 identified keywords, serving as pivotal focal points for research in this field. These keywords play a secondary role in aiding the comprehension of trends within the domain of quantum machine learning research. The utilization of VOSviewer software facilitated a network analysis based on keywords. This analysis involved conducting a co-occurrence analysis of all keywords, allowing the researcher to uncover the development and structure of the research field.

Notably, Machine Learning emerged as the most frequently occurring keyword, with a count of 470 in the literature spanning from 2000 to 2023. This was followed by Humans (249), Algorithms (174), Quantum Theory (153) and others. The co-occurrence relationship among these top 67 keywords is visually depicted in Figure \ref{fig:co-occurence-keywords}, where the size of each bubble corresponds to its frequency of occurrence and total strength of links with other keywords. In the keyword co-occurrence network, Machine Learning stands out as the node having largest diameter and font size, indicated by its significance in the research landscape. Nodes sharing the same color represent a single cluster, with Humans and Algorithms ranking as the second and third keywords, both belonging to the same color cluster. The links between nodes reflect the tendency of keywords to appear together in the same document. These 67 keywords were categorized into 6 clusters, each denoted by a distinct color. Cluster 1, highlighted in pink, encompasses 18 keywords, followed by cluster 2 (green) with 16 keywords, cluster 3 (blue) with 15 keywords, and clusters 4 (yellow), 5 (purple), and 6 (light blue) with 9, 8, and 1 keyword(s) respectively.

\subsection{Total Patent Citations by Year of Publications}

\begin{figure}[]
    \centering
    \includegraphics[width=0.8\textwidth,height=0.5\textwidth]{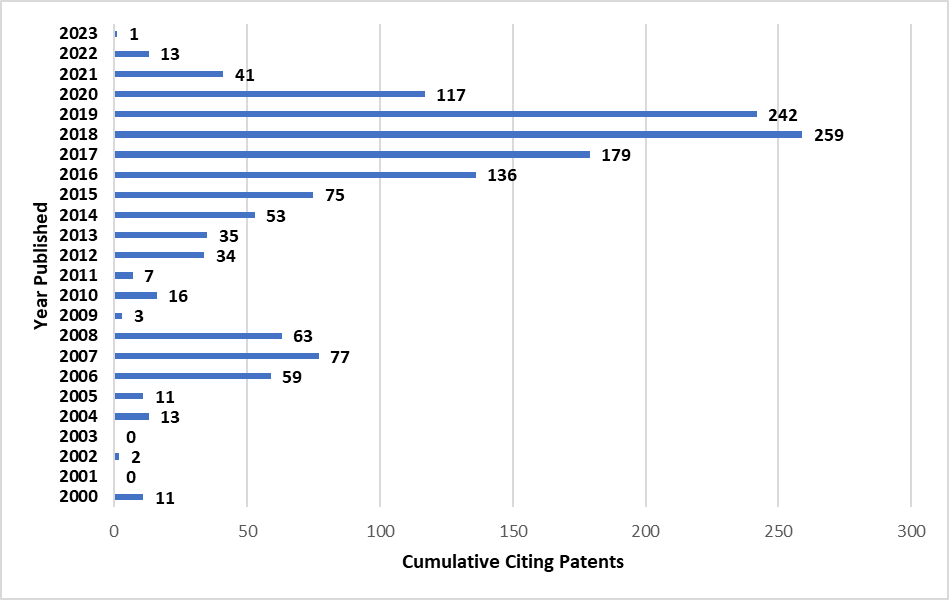}
    \caption{Total Patent Citations by Year of Publications}
    \label{fig:patent-per-year}
\end{figure}

The visual representation in Figure \ref{fig:patent-per-year} illustrates the annual count of patent citations. The collective count of citing patents for all scholarly works cited in patents within the result set was found to be 1447. Notably, there is a notable surge in the number of patent citations starting from 2014, indicating a period of heightened innovation and increased intellectual property activity in the domain of quantum machine learning. However, the data also suggests a subsequent decrease in patent citations from the year 2020 onward. The peak number of patent citations was identified in the year 2018, reaching the highest value of 259. This observed trend may prompt further investigation into the factors influencing the fluctuations in patent activity during these specific time periods, offering valuable insights into the dynamics of research and development in quantum machine learning.

\subsection{Top patent cited works over time}

In the field under investigation, a thorough examination of 9493 scholarly works has been conducted, revealing that 455 of these works have garnered attention by being cited in patents. The outcomes, presented in Table \ref{tab:patent-cited}, offer a detailed overview of the top 14 scholarly works with the highest patent citations, providing dynamic insights into the evolution of their impact in the realm of intellectual property.

Evidently, the article titled Simulation of networks of spiking neurons: A review of tools and strategies (\cite{brette2007simulation}) stands out at the forefront, accumulating the highest number of patent citations, totaling 50. In close pursuit is the article A lambic: a privacy-preserving recommender system for electronic commerce (\cite{aimeur2008alambic}) with 41 patent citations. Notably, the geographical distribution of these impactful articles reveals a concentration in the United Kingdom, where 9 of the top 14 articles originate which is more than the 50\% of the top patent cited works over the time. This is followed by contributions from the Netherlands (3), Germany (1), and the United States (1). These findings underscore the real-world applicability and influence of specific scholarly works, as evidenced by their integration into patents, thereby contributing significantly to practical advancements in the field.

\begin{table*}[]
\centering
\caption{Top patent cited works over time}
\scriptsize
  \begin{tabular}{llccl}
  \toprule
\textbf{} & \textbf{Title}                                                                                                                              & \textbf{Year} & \textbf{Citing Patents} & \textbf{Country} \\
\midrule
1 & Simulation of networks of spiking neurons: A review of tools and strategies                & 2007          & 50                                                                      & Netherlands      \\ 
2 & A lambic : a   privacy-preserving recommender system for electronic commerce                 & 2008          & 41                                                                      & Germany          \\ 
3 & Molecular   Graph Convolutions: Moving Beyond Fingerprints                                                                                  & 2016          & 30                                                                      & Netherlands      \\ 
4 & Timing,   Sequencing, and Quantum of Life Course Events: A Machine Learning Approach        & 2006          & 28                                                                      & Netherlands      \\ 
5 & Extending the lifetime of a quantum bit with error   correction in superconducting circuits & 2016          & 24                                                                      & United Kingdom   \\ 
6 & Barren plateaus in quantum neural network   training landscapes                                                                             & 2018          & 20                                                                      & United   Kingdom \\ 
7 & Decoding   DNA, RNA and peptides with quantum tunnelling                                                                                    & 2016          & 19                                                                      & United Kingdom   \\ 
8 & Nature -   Planning chemical syntheses with deep neural networks and symbolic AI          & 2018          & 19                                                                      & United   Kingdom \\ 
9 & Quantum   principal component analysis                                                                                                      & 2014          & 18                                                                      & United Kingdom   \\ 
10 & Quantum machine learning                                                                                                                    & 2017          & 16                                                                      & United   Kingdom \\ 
11 & Using   Quantum Confinement to Uniquely Identify Devices                                                                                    & 2015          & 16                                                                      & United Kingdom   \\ 
12 & The quantum internet                                                                                                                        & 2008          & 16                                                                      & United   Kingdom \\ 
13 & Quantum   Support Vector Machine for Big Data Classification                                                                                & 2014          & 15                                                                      & United States    \\ 
14 & Prediction   and real-time compensation of qubit decoherence via machine learning & 2017          & 15                                                                      & United Kingdom   \\ 
\bottomrule
\end{tabular}%
\label{tab:patent-cited}%
\end{table*}%

\subsection{Top Funding Organizations}

Table \ref{tab:funding-org} presents a detailed breakdown of countries and their corresponding funding contributions to Quantum Machine Learning (QML) research. Notably, The National Natural Science Foundation of China stands out as the leading funding agency, supporting 678 documents. Another Chinese funding agency (National Key Research and Development Program of China) is associated with 76 documents, contributing to China's overall dominance and significant investment in research and development. United States, as another major player, allocates its support across two funding agencies, resulting in a total of 432 documents.

\begin{table*}[]
\centering
\caption{Top Funding Organizations}
\scriptsize
  \begin{tabular}{lllccc}
  \toprule
\textbf{} & \textbf{Funding   Agencies}                                                                               & \textbf{Country} & \textbf{Document   Count} & \textbf{Citing Patents} & \textbf{Total Citations} \\ 
\midrule
1 & National   Natural Science Foundation of China                                                            & China            & 678                                                                     & 49                                                                      & 10301                                                                    \\ 
2 & National Science Foundation                                                                               & United   States  & 244                                                                     & 30                                                                      & 6670                                                                     \\ 
3 & U.S.   Department of Energy                                                                               & United States    & 188                                                                     & 19                                                                      & 5767                                                                     \\ 
4 & Deutsche Forschungsgemeinschaft                                                                           & Germany          & 129                                                                     & 17                                                                      & 4792                                                                     \\ 
5 & Engineering and Physical Sciences     Research Council        & United Kingdom   & 96                                                                      & 17                                                                      & 2879                                                                     \\ 
6 & H2020 European Research Council                                                                           & EU               & 78                                                                      & 3                                                                       & 3754                                                                     \\ 
7 & National Key Research and Development Program of China    & China            & 76                                                                      & 1                                                                       & 945                                                                      \\ 
8 & Japan Society for the Promotion of Science                                                                & Japan            & 74                                                                      & 1                                                                       & 1457                                                                     \\ 
9 & National   Research Foundation of Korea                                                                   & Korea            & 64                                                                      & 13                                                                      & 1162                                                                     \\ 
10 & Fundamental   Research Funds for the  Central Universities & EU               & 57                                                                      & 10                                                                      & 753                                                                      \\ 
\bottomrule
\end{tabular}%
\label{tab:funding-org}%
\end{table*}%

\begin{figure*}[!ht]
    \centering
    \includegraphics[width=1\textwidth,height=8cm]{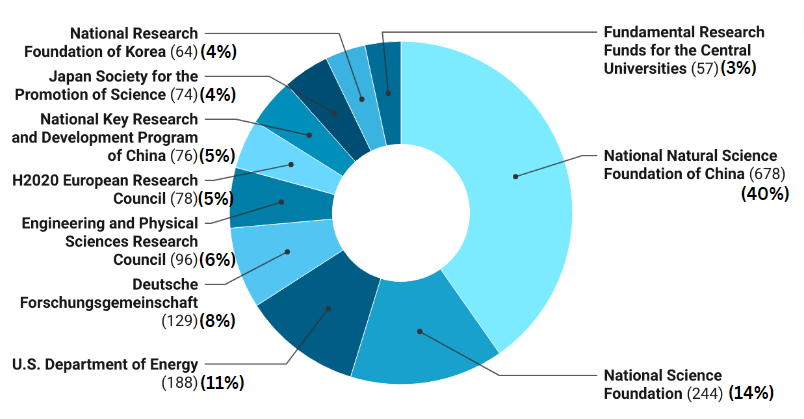}
    \caption{Top Funding Organizations}
    \label{fig:top-funding-org}
\end{figure*}

The accompanying donut chart (Figure \ref{fig:top-funding-org}) visually emphasizes the distribution of funding agencies across the countries. China emerges as the leading investor, contributing to almost 45\% (40\% of National Natural Science Foundation of China and 5\% of National Key Research and Development Program of China) of the total funded projects. United States follows closely, accounting for nearly 25\% (14\% of National Science Foundation and 11\% of U.S. Department of Energy) of the total projects. The combined efforts of these two major contributors constitute approximately 70\% of the total project, reflecting a significant concentration of funding influence. Additionally, other nations collectively contribute to the remaining 30\% of funded projects. This data underscores the global nature of QML research funding, with a notable concentration in key countries.

\section{Conclusion} \label{conclusion}

This paper conducts a bibliometric analysis of Quantum Machine Learning spanning the years 2000 to 2023, utilizing the Lens database. Despite the growing significance of Quantum Machine Learning in both academia and industry, the global literature in this field remains relatively small, comprising only 9493 publications over a 24-year period from 2000 to 2023. Employing VOSviewer, Excel, the analysis explores publication characteristics, influential countries and institutions, key authors, patent-cited works, funding organizations, journals, and science mapping.

The number of publications has exhibited a nearly steady increase since 2015, indicating a growing interest in Quantum Machine Learning. The United States emerges as the most productive country, with the Chinese Academy of Sciences standing out as the most influential institution. The analysis underscores China's prominent role as a leading investor in this field. Noteworthy authors include O Anatole von Lilienfeld, contributing the highest number of publications, and Masoud Mohseni, with the highest number of citations.

In conclusion, Quantum Machine Learning has the potential to revolutionize various sectors, expanding the capabilities of computers and data analysis. Despite being in its early stages, ongoing advancements in quantum technology and the evolution of quantum-inspired algorithms present promising possibilities for the future.

\section{Limitations} \label{limitations}
Despite our effort to provide a comprehensive review, it's crucial to acknowledge the dynamic nature of the literature in this field. Numerous studies may have been omitted due to the rapidly evolving nature of the research. This study relies solely on Lens database for analysis, and future researchers are encouraged to broaden their scope by incorporating articles from diverse databases such as Web of Science, Scopus and Google Scholar, ensuring a more inclusive representation of the available literature. Also this study include a focus on a singular document type (journal articles) without incorporating a broader range of sources such as books, book chapters, conference articles, among others. Furthermore, the search parameters employed in this study may not capture the entire spectrum of relevant articles. To enhance the comprehensiveness of future analyses, researchers could explore additional keywords, refining the search strategy for a more exhaustive review.

In summary, while this study contributes valuable insights, there are inherent limitations that future researchers should address to further enrich our understanding of the dynamic and expanding field under investigation.

\bibliographystyle{unsrt}
\bibliography{references}  






\end{document}